# Which cities produce more excellent papers than can be expected?

# A new mapping approach—using Google Maps—based on statistical significance testing

*Journal of the American Society for Information Science and Technology* (forthcoming)

Lutz Bornmann [$] and Loet Leydesdorff [§]

[$] Max Planck Society, Hofgartenstr. 8, 80539 Munich, Germany; bornmann@gv.mpg.de.

[§] Amsterdam School of Communications Research, University of Amsterdam, Kloveniersburgwal 48, NL-1012 CX Amsterdam, The Netherlands; loet@leydesdorff.net; www.leydesdorff.net.

Corresponding author: Lutz Bornmann, bornmann@gv.mpg.de.

**Abstract**
The methods presented in this paper allow for a statistical analysis revealing centers of excellence around the world using programs that are freely available. Based on Web of Science data (a fee-based data base), field-specific excellence can be identified in cities where highly-cited papers were published more frequently than can be expected. Compared to the mapping approaches published hitherto, our approach is more analytically oriented by allowing the assessment of an observed number of excellent papers for a city against the expected number. Top performers in output are cities in which authors are located who publish a statistically significant higher number of highly-cited papers than can be expected for these cities. As sample data for physics, chemistry, and psychology show, these cities do not necessarily have a high output of highly-cited papers.

**Key words**
Scientific excellence; Highly-cited papers; Geography of science; Overlay; Spatial scientometrics; Google Map

# 1 Introduction

How can scientific excellence be evaluated in geographic regions and cities? In this study, we explore new methods to analyze the geographic distribution of highly-cited papers. We build, on the one side, on previous studies in the geography of science; these were recently reviewed by Frenken et al. (2009). This review listed mainly descriptive studies which investigated differences among regions or countries, respectively, in terms of their publication output and/or citation counts. On the other side, the methods proposed in this study add to our former research activities in the area of spatial bibliometrics. Leydesdorff and Persson (2010) used Google Earth, Google Maps, and/or network visualization programs to overlay the network of relations among addresses in scientific publications onto the geographic map. Bornmann, Leydesdorff, Walch-Solimena, and Ettl (in press) drew on this publication and presented methods to map centers of excellence around the world using Scopus data and programs that are freely available. These programs allow for the identification of field-specific excellence and their agglomerations in cities. Bornmann and Waltman (in press) used additional visualization methods (density maps) for a spatial examination revealing regions of excellence (see here also van Eck & Waltman, 2010). In contrast to the other approaches which focus on cities, this latter method deals with broader regions where highly-cited papers were published.

One advantage of the approaches for the spatial visualization of concentrations of highly-cited papers published hitherto has been the relatively straightforward identification of cities and regions with a high output in this respect. However, we consider it as a disadvantage that only the quantitative numbers of highly cited papers are then used. If a city has a very high paper output in general (that means not only of highly cited papers) one can expect a high number of excellent (that means highly-cited) papers proportionally (see here Bornmann, Mutz, Marx,

Schier, & Daniel, 2011; Bornmann, Schier, Marx, & Daniel, 2011). For example, if authors located in one city have published 10,000 papers, one would expect for statistical reasons that approximately thousand (that is, 10%) would also belong to the top-10% most-highly cited papers. An observed number of 700 highly-cited papers for this city may seem as a large number compared to other cities, but it turns out to be smaller than one would expect in this case. Thus, the number of highly-cited papers for a city should be assessed statistically given the number of publications in total. This approach of comparing observed with expected numbers of highly cited papers is one method for measuring scientific excellence besides others (e.g., Costas, Bordons, van Leeuwen, & van Raan, 2009; Seglen & Aksnes, 2000).

In this study, we propose a procedure which considers the total numbers of papers and the most-highly cited ones in order to assess the status of a city. The procedure can be run by using Web of Science (WoS, a fee-based data base) data and programs which are freely available on the Internet and easy to handle (http://www.leydesdorff.net/topcity). The results can be visualized on a Google Map and thus show for each city with at least a single excellent paper how large the difference between observed and expected highly-cited papers is. As examples, we use samples of papers published in 2008 in physics, chemistry, and psychology. Consequently, the window for gathering citations to these papers amounts to approximately three years: from 2008 until the date of data collection in February 2011, respectively. Glänzel, Thijs, Schubert, and Debackere (2009) argued that the "use of a three-year citation window suffices for building both relative and high-impact citation indicators" (p. 186).

From a policy perspective, it may pay off for the sciences within a country to identify (by means of these visualization methods) and expand regional centers of excellence (for example, with specific financial support). In our opinion, one should not subsidize size, but those centers should be fostered where the observed number of excellent papers exceeds the expected numbers. Cities are a relevant level for the assessment, since a high probability of co-operation among scientists working at relatively short physical distances is often mentioned in the literature. Synergies between ideas in direct face-to-face communication between scientists are held to be major factors stimulating productivity (Matthiessen, Schwarz, & Find, 2002; Wagner, 2008).

## 2 Methods

*Procedure to generate the underlying data*

Tijssen, Visser, and van Leeuwen (2002) and Tijssen and van Leeuwen (2006) argued that the top-10% of papers with the highest citation counts in a publication set can be considered as highly cited (see also Lewison, Thornicroft, Szmukler, & Tansella, 2007). In this study we follow this classification and focus on the top-10% of papers published in 2008 in the three fields under study, using a citation window from this year of publication (2008) up to the date of harvesting data from the WoS for this research (February 2011). Since this paper is intended to introduce our new mapping approach, we used only a sample of all papers published in each field. Sets of approximately 10,000—containing approximately 1,000 papers in the top-10%—are sufficiently large for our purpose.

In the following, the procedure to map the cities of the authors having published the top-10% most-highly-cited papers in a certain field is described in detail so that one is able to reproduce these methods. We explain the procedure for the field of "chemistry" (date of search: February 2011). With the search string "PY=(2008) AND Document Type=(Article). Timespan=All Years. Databases=SCI-EXPANDED" in the advanced search field of WoS (version 4.1) all papers with the document type "article" were retrieved which had been published in 2008. The timespan "all years" was used because papers published in 2008 may have been entered into the database in other years. We restricted the search to articles (as document types) since (1) the method proposed here is intended to identify excellence at the research front and (2) different document types have different expected citation rates, possibly resulting in non-comparable datasets.

This search results indicate that more than 100,000 papers can be retrieved. To refine the results to a single field, "more options / values..." in "Subject Areas" of the frame "Refine Results" can be clicked. When all subject areas (a subject area is also called a journal set or a cluster of similar field-specific journals) have been made visible on the screen, the subject areas can be sorted alphabetically. For the field chemistry, let us select all subject areas which begin with the word "Chemistry" as follows:

CHEMISTRY, ANALYTICAL (1,227 articles)

CHEMISTRY, APPLIED (870 articles)

CHEMISTRY, INORGANIC & NUCLEAR (670 articles)

CHEMISTRY, MEDICINAL (560 articles)

CHEMISTRY, MULTIDISCIPLINARY (4,513 articles)

CHEMISTRY, ORGANIC (1,349 articles)

CHEMISTRY, PHYSICAL (2,418 articles)

The selection of these chemistry categories results conveniently in 10,460 articles in total. A search in WoS (version 5.2) for all chemistry papers published in 2008 shows that this number constitutes a sample of nearly 10% of all (117,411)[1] chemistry articles published in this year. This (and also the other drawn samples for this study) cannot be considered as a random sample from the entire population or a representative sample. In all likelihood, it consists of papers published rather at the end than at the beginning of 2008.

Note that the number of 10,460 chemistry articles is different from the sum of the articles for the single categories (e.g., CHEMISTRY APPLIED) since some articles are categorized by Thomson Reuters in more than a single subfield of chemistry. The 10,460 papers ("full records")[2] can be saved in packages of 500 articles each as plain text (e.g., savedrecs500.txt). The resulting 21 packages are then merged into a single file "data.txt" (see here the instructions on http://www.leydesdorff.net/software/isi/index.htm). This file is stored on the disk in a separate folder "All papers."

The following procedure should be followed.[3] The programs cities1.exe and cities2.exe are copied from the website into the folder. These programs (including the respective user instructions) can be downloaded from http://www.leydesdorff.net/maps (see here Leydesdorff & Persson, 2010). Upon running, cities1.exe will prompt the user with the question: "Do you wish to skip the database management?" This question should in this case be answered with "N" (meaning: no). Later on, four questions follow: with the first and second questions one can set a threshold in terms of a minimal percentage of the total set of city-names in the data or set a minimum number of occurrences. The default answers to the questions ("0") can all be accepted.

---

[1] Although the new version 5 of the WoS provides this number, the download is equally restricted to 100,000 records. We used version 4 because the address field in version 5 was not yet in good shape at the date of this research.
[2] For our purpose, it is not necessary to download abstracts or cited references. This shortens the processing time both on the Internet and locally.
[3] A more detailed instruction including examples can be found at http://www.leydesdorff.net/topcity/index.htm.

The third and fourth questions enable the user to obtain a cosine-normalized data matrix and to generate network data. Both questions can for our purpose be answered with "N" (meaning: no).

The program cities1.exe creates among other files the file named cities.txt. This file contains all city entries from data.txt, but organized so that this data can be "geo-coded," that is, provided with latitudes and longitudes on a map. If more than a single co-author of a publication with an identical address is included, this leads to a single address (or a single city occurrence) in cities.txt. If the scientists are affiliated with different departments within the same institution, this leads to at least two addresses or two city names, respectively. The counting of occurrences in this study is the result of the procedure of how authors' addresses on publications are gathered by Thomson Reuters for inclusion in the WoS. Thus, we use so-called "integer counting".[4]

The content of cities.txt can be copied-and-pasted into the GPS encoder at http://www.gpsvisualizer.com/geocoder/ (using Google as source for the geocoding).[5] Since no more than 1,000 entries can be processed by the encoder, larger numbers of lines than 1,000 entries in cities.txt will have to be entered into the encoder in subsequent steps. After saving the results in the output window of the geo-encoder as a DOS text file (e.g., "geo.txt") this data can serve as input to cities2.exe. If geo.txt contains all entries from cities.txt in the same order but with the additional geo data, the program cities2.exe can be used for matching these files. This program first prompts for the name of the input file (in this case: "geo.txt") and then produces a number of output files in various formats within the same folder, among which the file

---

[4] Integer counting is distinguished from fractional counting in which case each address or author is attributed with a proportional weight.
[5] The default geocoder at this website is Yahoo!. Both Yahoo! and Google may miss addresses or place them erroneously. We found Google somewhat more reliable, but not when using Asian addresses. Considerable effort was made to produce a file "cities.txt" containing additional information where necessary. For example, in the case of US cities, state abbreviations are included in order to be able to distinguish, for example, between "Athens, OH" and "Athens, GA." One can additionally consider to fill out the addresses listed as "0,0" for the coordinates in the resulting file using the other engine.

"inp_gps.txt". This file is in the format so that it can be used for the generation of overlays to Google Maps and Google Earth at http://www.gpsvisualizer.com/map_input?form=data.

In a final step, we proceed with the statistical procedure. Topcity2.exe (available at http://www.leydesdorff.net/topcity/topcity2.exe) first asks to specify a percentile level. In this study, we used the top-10% of the most cited papers, and accordingly ten percent (the default) was entered. However, the user can vary this parameter. The file "ztest.txt" is one output file of topcity2.exe, and can be uploaded into the GPS Visualizer at http://www.gpsvisualizer.com/map_input?form=data. The data relevant for statistical analysis are provided in the file ucities.dbf.

The webpage of the GPS Visualizer offers a number of parameters that can be set to visualize the information contained in "ztest.txt." We suggest to change the following parameters: change (a) "waypoints" into "default;" (b) "colorize using this field" into "custom field" and choose "color" in this field; (c) "resize using this field" into "custom field" (d) in "custom resizing field" "n" is written and (e) at "Maximum radius" replace 16 with 30.

After processing the GPS data, the Google map is displayed first in a small frame, but this map is also available as full screen. The map shows the regional distribution of the authors of highly cited papers (cities with authors who published at least one excellent paper in the sample). The opacity of the background map can be adjusted and other layouts are also available in Google or Yahoo!. With the instruments visualized on the left side of the map one can zoom into the map. (Initially, the global map is shown.) For the maps presented here below we zoomed into Europe in order to generate comparable maps for different publication fields. However, other regional foci can also be chosen.

In order to determine the quotient of observed and expected numbers of excellent papers for a specific city, one can click on the respective city. The number is then displayed in the respective labels. Maps generated in this way can be copied to other programs (like Microsoft Word) by using programs utilized for screen shots (e.g., using the PrtScr key, Screengrab! in Firefox, or another program such as Hardcopy). If one uses the download instead of the view command shown in the Google Maps output page, a html-coded page can be saved that includes the data. Opening this page within a browser will regenerate the respective Google Map. Additionally, one can ask Google Maps for an API (free) and upload the file at one's homepage.

Two problems are inherent to the approach proposed here. The user should always be aware of the limitations when this methodology is applied. (1) The methods as described above do not allow for the identification of research institutions on the map where the authors of the excellent papers are located. (2) Two effects could be responsible for a high number of publications visualized on the map for one single city: (a) many scientists located in this city (i.e., scientists at different institutions or departments within one institution) produced at least one excellent paper or (b) one or only a few scientists located in this city produced many influential papers. With the approach in this study – assuming cities as units of analysis – one is not able to distinguish between these two interpretations (see here Bornmann, et al., in press; Bornmann & Waltman, in press). In general, this method does not inform us about the social processes underlying the generation of highly-cited papers.

We advise to check the maps against the original data at a number of random places before exhibiting it on the web. The user has all statistical data available in the file ucities.dbf which is created by topcity2.exe and can be opened using Excel or SPSS. In the case of systemic error, we appreciate and will respond to feedback.

*Statistical procedure*

The *z* test for two independent proportions is used for evaluating the degree to which an observed number of top-cited papers for a city differs from the value that would be expected on the basis of the null hypothesis of randomness in the selection of papers for a city (Sheskin, 2007, pp. 637-643). If excellent papers are defined as the top-10% most-highly cited papers in a field, on the basis of this null hypothesis one could *expect* 10% of all papers published in each city, respectively, to belong to this category. Ten percent of the papers can be expected to belong to the 10% most-highly cited papers in a certain field.

The following equation is employed to compute the test statistic:

$$z = \frac{p_o - p_e}{\sqrt{p(1-p)(2/n)}}$$

where: *n* represents the number of all papers published by authors located at this city;

$p_o = n_o/n$ represents the proportion of papers (published by authors located at this city) that are among the top-10% most-highly cited papers ($n_o$ represents the number of observed papers in this category);

$p_e = 0.10$ is the proportion of papers expected to be among the top-10% most-highly cited papers;

$$p = \frac{n_o p_o + n_e p_e}{2n}$$

where: $n_e = n/10 =$ represents the expected number of the top-10% most-highly cited papers.

*z* is positively signed if the observed number of top papers is larger than the expected number and negatively signed in the reverse case. Since the *z* test for two independent proportions is a "large sample procedure for evaluating a 2 x 2 contingency table" (Sheskin, 2007, p. 637), $n_e$ should be equal to or larger than 5. An absolute value of $z$ larger than 1.96 indicates statistical significance at the five percent level ($p<.05$) for the difference between observed and expected numbers of top-cited papers. In other words, the authors located at this city are outperformers with respect to scientific excellence in terms of this statistics.

Using this statistical procedure, we designed the city circles which are visualized on the map using different colours and sizes. The radii of the circles are calculated by using:

| observed value – expected value | + 1

The "+1" must prevent the circles from disappearing if the observed ratio is equal to the expected one.

Furthermore, the circles are coloured green if the observed values are larger than the expected values. We use dark green if both the expected value is at least five (and a statistical significance test is legitimate) and $z$ is statistically significant; light green indicates a positive, but statistically non-significant result. The in-between colour of lime green is used if the expected value is smaller than five and a statistical significance test hence should not be calculated. One should be cautious with interpretations of results below this threshold value.

In the reverse case that the observed values are smaller than the expected values the circles are red or orange, respectively. They are red if the observed value is significantly smaller than the expected value and orange-red if the difference is statistically non-significant. If the requirement for the test of an expected value larger than five is not fulfilled the circle is coloured orange. If the expected value equals the observed value a circle is coloured grey.

# 3    Results

In order to show the fruitfulness of the proposed method we provide three field-specific maps based on sample data. Figure 1 shows the location of authors in Europe having published highly-cited papers in physics. The map is based on a sample of the top-10% of articles published in 2008 in a WoS journal set for physics. The physics sample consists of 9,950 articles (searched in WoS, version 4.1, on February 2, 2011) categorized by Thomson Reuters into the following journal sets:

PHYSICS, APPLIED (3,427 articles)

PHYSICS, ATOMIC, MOLECULAR & CHEMICAL (905 articles)

PHYSICS, CONDENSED MATTER (2,578 articles)

PHYSICS, FLUIDS & PLASMAS (522 articles)

PHYSICS, MATHEMATICAL (996 articles)

PHYSICS, MULTIDISCIPLINARY (2,425 articles)

PHYSICS, PARTICLES & FIELDS (772 articles)

The number of 9,950 is different from the sum of the articles for the single categories since some articles are categorized in more than a single subfield. A search in WoS version 5.2 shows that this sample (n=9,950) is nearly 10% of all (117,244) papers published in the physics category in 2008. The top-10% most-highly-cited articles in the sample are those 1,071 papers which received at least eight citations each between their publication in 2008 and the date of the search.

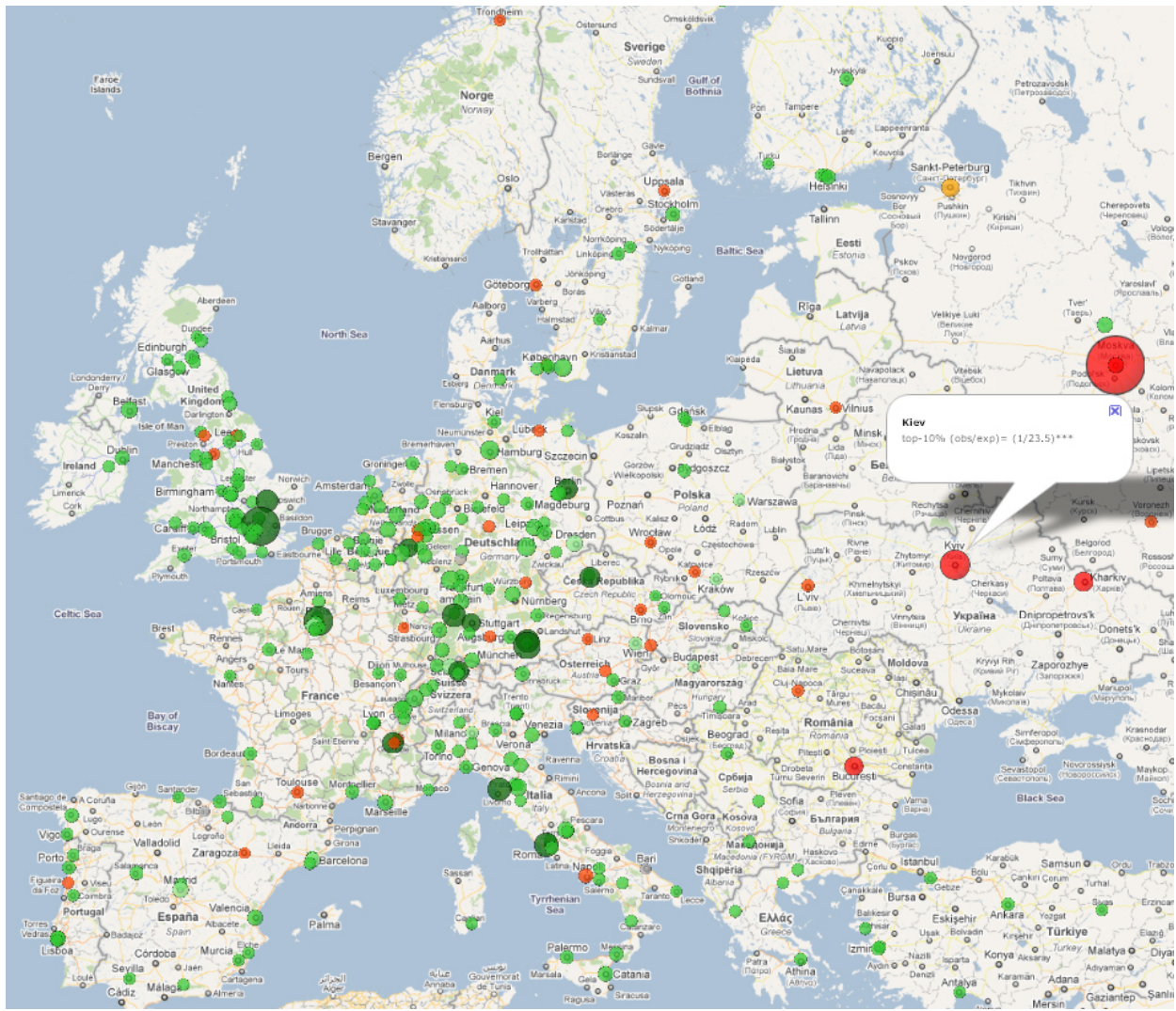

**Figure 1**. Cities in Europe in which highly cited papers in physics were published in 2008. The global map is available at http://www.leydesdorff.net/topcity/figure1.htm.

Figure 1 shows the cities of authors of the excellent papers (the circles with different colors on the map) and the deviations of the observed from the expected number of top-cited papers per location (the circle radii). If one clicks on a circle (at http://www.leydesdorff.net/figure1.htm), a frame opens showing the number of observed versus expected values for the respective city, as well as an asterisk (*) indicating whether the difference between the values is statistically significant or not at the five-percent level.

For example, the label for Kiev shows the observed versus expected numbers for this city. The city is indicated by a red circle because of a larger expected than observed value. In the frame the large and statistically significant difference between observed (n=1) and expected value (n=23.5) in the sample is provided as a ratio. The largest red circle on this map is indicated for Moscow. Although Moscow has a high output of papers in physics, the number of top-cited papers is comparably small in the sample. Large green circles on the map are visible for London, Paris, Karlsruhe, Munich (and Garching), Pisa, and Rome. The largest difference between observed and expected values is for London; 33% of all published papers belong to the top-10% most-highly-cited papers. Cambridge in the proximity of London also has a dark green circle. The observed citation counts are higher than the expected counts and this difference is statistically significant.

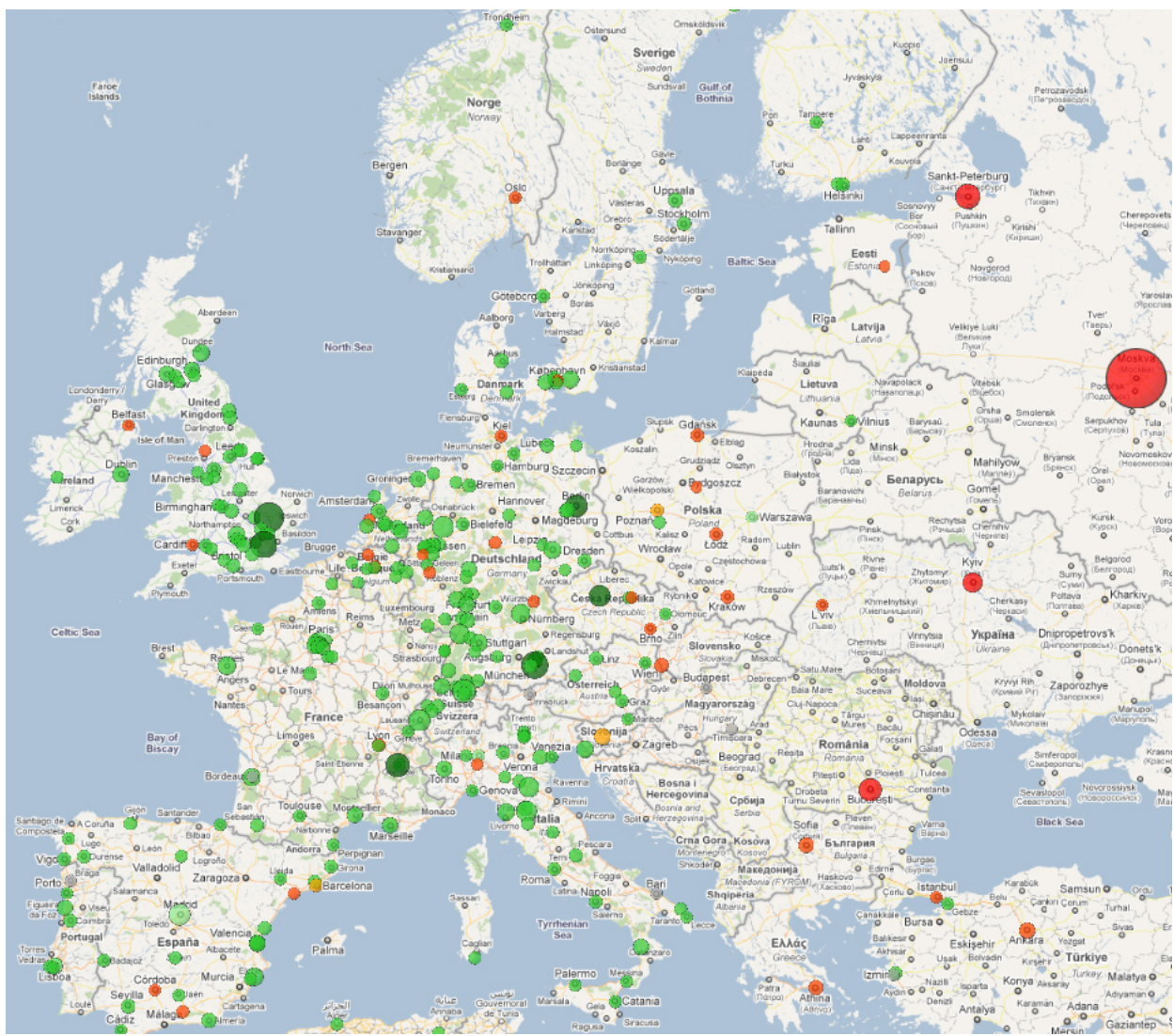

**Figure 2**. Cities in Europe in which highly cited papers in chemistry were published in 2008. The global map is available at http://www.leydesdorff.net/topcity/figure2.htm.

Figure 2 shows the corresponding map for the chemistry sample data (specified above in section 2). For chemistry, one can see a similar distribution of red and green circles as for the physics sample data (see Figure 1). The more one moves to the east of Europe the fewer green circles are visible. High densities of green circles can in this case be found in south-western

Germany, the south-eastern part of UK, as well as areas in Denmark (around Copenhagen), the Netherlands, Switzerland, and Italy. As in the physics map a large red circle is positioned at Moscow. Although in the sample five top-cited papers were published from authors located here, compared to the expected value, this observed value is low. In agreement to physics, a higher number of the observed against the expected value is indicated for Cambridge in this case. Prague also has a higher than expected contribution to the top-10% layer in the sample data for chemistry.

Since physics and chemistry are both natural science fields it could be interesting to see how the European map changes in the case of psychology. Figure 3 shows the map for the authors of the top cited papers. All in all, 1,592 top cited papers were selected in the Social Science Citation Index (WoS, 4.1) on the basis of 15,142 articles published in 2008. This represents more than half of the entire population of (24,877) papers within this category (as searched in WoS, version 5.2):

PSYCHOLOGY (2,398)

PSYCHOLOGY, APPLIED (1,400)

PSYCHOLOGY, BIOLOGICAL (591)

PSYCHOLOGY, CLINICAL (2,988)

PSYCHOLOGY, DEVELOPMENTAL (1,954)

PSYCHOLOGY, EDUCATIONAL (991)

PSYCHOLOGY, EXPERIMENTAL (3,022)

PSYCHOLOGY, MATHEMATICAL (444)

PSYCHOLOGY, MULTIDISCIPLINARY (3,436)

PSYCHOLOGY, SOCIAL (1,670)

Each of the 1,592 top cited papers in the sample received at least eight citations between 2008 and the date of research (March 2, 2011). Ten percent of 15,142 is 1,514. However, topcity2.exe includes ranks which are tied at the percentile level chosen by the researcher and thus we retrieved 1,592 records using the 10% level.

In Figure 3, it is interesting to see a much smaller circle for Moscow than in both other figures. In the psychology sample data, the total output with an address in Moscow is small. The largest set in Europe (in this psychology sample) is attributed to London: 147 top-cited papers against 71.50 expected papers. Thus "London, UK" outperforms the expectation to a statistically significant degree. However, with 45 papers in the top-10% (against 19.4 expected), Berlin outperforms London in this sample with an observed/expected rate of 2.32 ($p < 0.05$) against 2.06 ($p < 0.05$) despite the lower number in the sample. The comparison of London and Berlin points out the importance of considering not only the real numbers of excellent papers but also the expected numbers.

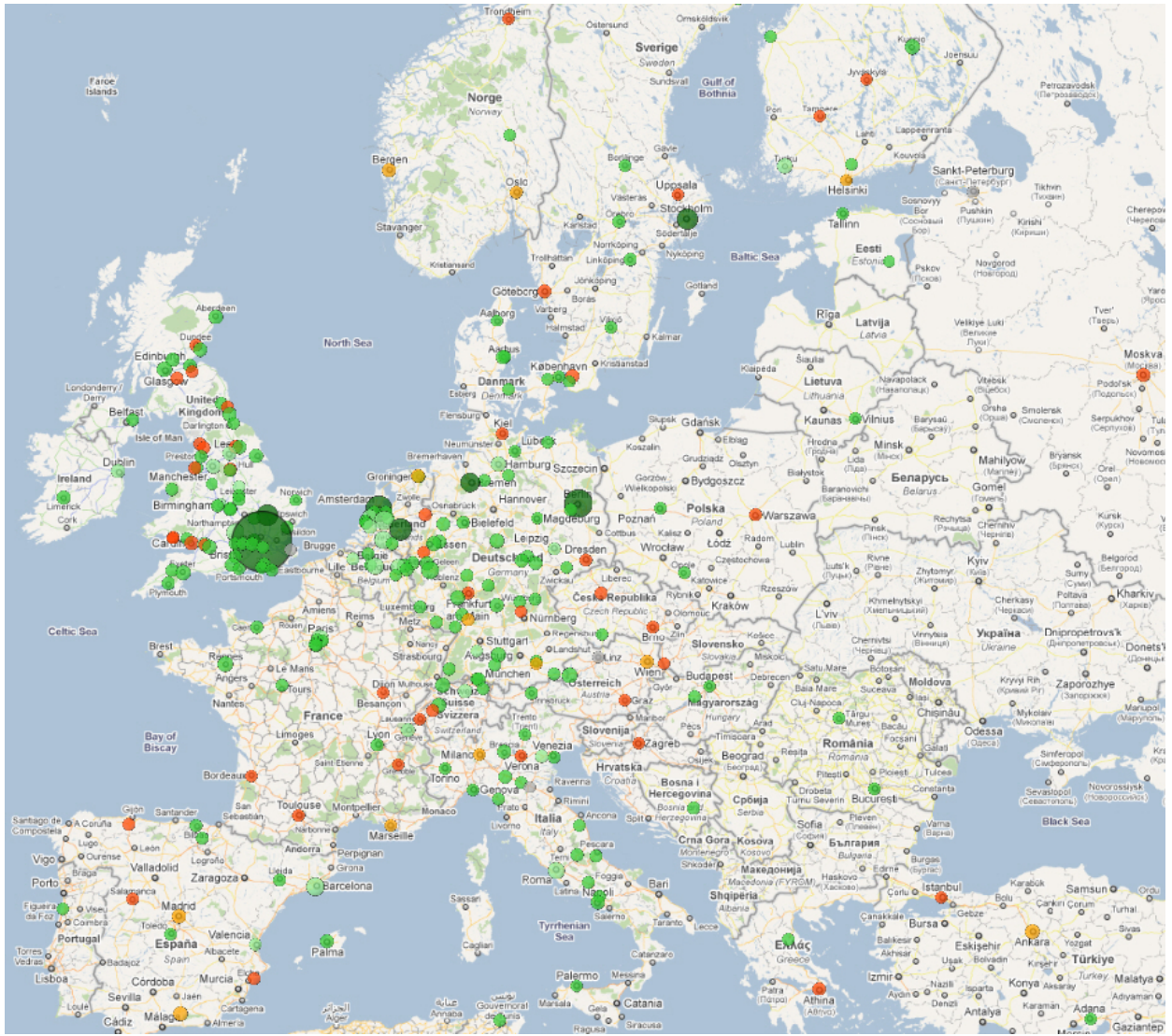

**Figure 3**. Cities in Europe in which highly cited papers in psychology were published in 2008. The global map is available at http://www.leydesdorff.net/topcity/figure3.htm.

## 4  Discussion

A number of works have been published on productivity rankings and centres of excellence (see, e.g., Buela-Casal, Gutiérrez-Martínez, Bermúdez-Sánchez, & Vadillo-Muñoz, 2007; Costas, et al., 2009; Frenken, et al., 2009; Glänzel, et al., 2009). The methods presented in this paper allow for an analysis revealing centers of excellence around the world using programs that are freely available. Based on WoS data, field-specific excellence can be identified in cities where highly-cited papers were published. Compared to the mapping approaches published hitherto, our approach is more analytically oriented by allowing the assessment of an observed number of excellent papers for a city against the expected number. With this feature, this approach can not only identify the top performers in output but the "true jewels." These are cities in which authors are located who publish significantly more top-cited papers than can be statistically expected. As the sample data sets in this paper show for three fields, these cities do not necessarily have a high output of highly-cited papers. Thus, our approach normalizes for size.

Despite the advantages of our approach by mapping observed versus expected numbers, we recognize the limitations inherent to these bibliometric exercises: (1) Publications are among several types of scientific activities. (2) It is not always the case that the addresses listed on the publication reflect the locations where the reported research was conducted. (3) The handling of multiple authorships is different in scientific fields. For example, the first and last authors have a special meaning in many, but not in all fields of science. Note that our method does not consider the different positions of authors on a paper. A paper is counted for a city if at least one co-author is located there – independently of whether this co-author is on the first or $26^{th}$ position. (4) No standard technique exists for the subject classification of articles. In bibliometrics, there is an ongoing debate about the proper classification (see here Bornmann, et al., in press; Leydesdorff & Rafols, 2009; Rafols & Leydesdorff, 2009). (5) Citation counts (and thus the categorization of publications as highly cited) are a function of many variables besides scientific quality (Bornmann & Daniel, 2008). (6) The ratios of observed versus expected values for the cities should be interpreted against the backdrop of the total number of excellent papers. This data is available in the file ucities.dbf for every city. From the policy perspective those cities are perhaps of interest only which have published a certain minimum number of excellent papers. As noted, if the expected value is below five and thus the number of papers below 50, significance testing should not be undertaken and this is indicated using another colour on the map.

Although we controlled for some of these other sources of variance in our study by restricting the analyses to single fields (operationalized as ISI Subject Categories), a single publication year, and a single document type, other factors can also have an influence. For example, it is clearly visible on the maps that Eastern Europe is dominated by red circles flagging a lower number of observed top-cited papers than can be expected. Marx (2011) and Cardona and Marx (2006) showed in case studies (e.g., the works of Vitaly Ginzburg) that errors in WoS can result in differences between the real number of citations for Russian papers and the number indicated as "Times Cited." For example, the top paper of Vitaly Ginzburg have "garnered 511 instead of 358 citations as stated in the WoS source item" (Cardona & Marx, 2006, p. 463). Although older Russian works might be more affected by these errors than younger (like those included in this study), the errors could result in systematic lower impact values for Russian papers (and thus a lower chance of being among the top-10% in a field).

In addition to the limitations that are inherent to using bibliometric data, we want to mention some limitations that are inherent to this methodology: The test for statistically significant differences between observed and expected values is only possible for a part of all cities (those with expected values larger than or equal to five; in other words, with total numbers $\geq 50$). In our physics data set, for example, only 90 of the 658 city addresses with at least one top-10% cited paper could thus be tested. Thus, the results for the cities in one field cannot directly be compared to each other. The use of larger set can ease this condition in the significance test.

A technical limitation of the statistics concerns the large number of pair-wise comparisons of observed versus expected values which are performed for cities of a single field (in physics, for example, 90 comparisons). In case of multiple significance tests, a Bonferroni correction of the significance level ($p < 0.05$) can be considered: for $n$ statistical tests each individual comparison is then tested at a statistical significance level of $1/n$ times (in case of physics: $p = (0.05/90 =) 0.0006$). Two considerations led us to disregard this Bonferroni correction: (1) the correction is conservative by tendency. In other words, it makes it difficult to reject the null hypothesis of no differences between observed and expected values. (2) Technically, it is not feasible to include the tables of various critical values for significance testing given the number of pair-wise comparisons within the routines without overloading the program. The advanced user has all statistical data available in the file ucities.dbf (which is generated by topcity2.exe) and can perform Bonferroni correction using programs such as SPSS.

Despite these limitations our proposed approach for the mapping of excellent papers provides many advantages against the approaches published hitherto. It should be the task of future research to develop the methodology further into a procedure without these limitations. The scientific mapping of excellent papers can complement the popular institutional rankings published so far. The Academic Ranking of World Universities (http://www.arwu.org/) published by the Center for World-Class Universities and the Institute of Higher Education of Shanghai Jiao Tong University (China) includes highly-cited researchers besides alumni and staff winning Nobel Prizes and Fields Medals and papers indexed in the Science Citation Index-Expanded and the Social Science Citation Index (see here Billaut, Bouyssou, & Vincke, 2010;

Dehon, McCathie, & Verardi, 2010; Jeremic, Bulajic, Martic, & Radojicic, 2011). However, none of these rankings hitherto test for the statistical significance of observed differences in performance.